\magnification=\magstep1
\hfuzz=6pt
\baselineskip=16pt

$ $

\vfill

\centerline{\bf The universal path integral}

\vskip 1cm

\centerline{Seth Lloyd$^1$\footnote{*}{Corresponding author: slloyd@mit.edu}, 
Olaf Dreyer$^2$}

\bigskip

\centerline{1. Department of Mechanical Engineering}

\centerline{Massachusetts Institute of Technology}

\centerline{Cambridge MA 02139 USA}

\bigskip

\centerline{2. Dipartimento di Fisica, Universit\`a di Roma `La Sapienza'}

\centerline{and Sez. Roma1 INFN, P.le A. Moro 2, 00185 Roma, Italy}

\vskip 1cm

\noindent{\it Abstract:} 
Path integrals represent a powerful route to quantization:
they calculate probabilities by summing over classical configurations
of variables such as fields, assigning each configuration a phase
equal to the action of that configuration.    This paper defines 
a universal path integral, which sums over all computable structures.
This path integral contains as sub-integrals all possible 
computable path integrals, 
including those of field theory, the standard model of elementary
particles, discrete models of quantum gravity, string theory, etc.
The universal path integral possesses a well-defined measure
that guarantees its finiteness, together with a method for extracting
probabilities for observable quantities.  The universal
path integral supports a quantum theory of the universe
in which the world that we see around us arises out of
the interference between all computable structures.

\vfill\eject

Over the past few decades it has become clear that physicists must
contemplate seriously the possibility that the observed laws of physics
may only represent one out of a large number of
possible laws.  In string theory, for example, the string
theory landscape identifies some $10^{500}$ possible vacua,
each of which could give rise to different properties for 
the laws of elementary particles, and only one of which might
correspond to our observed laws [1].  If our laws are plucked
from an ensemble of possible laws, it makes sense
to investigate from what types of ensembles of laws ours might arise,
and how one particular set of laws is plucked from the ensemble.  
To this end, this paper
proposes a universal path integral that encompasses all
computable path integrals.  The path integral possesses a
well-defined measure that makes finite all path integrals derived
from it, and provides a natural method for the prediction of 
observable probabilities.  The universal path integral provides
a mathematically well-defined framework for describing the
ensemble of computable quantum theories, and for determining
how the laws of nature as we observe them arise from this ensemble.

The ingredients of a path integral are a set of classical
configurations or `paths' $\{ \Phi \}$, for example, a set of configurations
for classical fields over spacetime, together with an action
$S( \Phi )$, a simply-computable function of the classical
configurations [2-3].  To perform the integral, one integrates
over all configurations that are compatible with the conditions $\Gamma$
that one wants to observe, e.g., configurations of fields
on initial and final spacelike surfaces, yielding an amplitude
for those conditions:
$${\cal A}(\Gamma) = \int_{\Phi \in \Gamma} {\cal D} \Phi ~ e^{iS(\Phi)}
.\eqno(1)$$
The probability of $\Gamma$ is then proportional to
$|{\cal A}(\Gamma)|^2$.  There are a number of technical issues
associated with evaluating such path integrals.  First of all,
one must define the measure of integration ${\cal D}\Phi$ in
a sensible way. 
Secondly, one must actually perform the integral: the integral
itself is highly oscillatory and hard to approximate. 
Finally, in the case of theories such as string theory or
eternal inflation that attempt
to provide a quantum-mechanical description of the universe as
a whole, the issue arises of how to extract probabilities for
observed features of the universe out of the path integral [4].

The universal path integral provides a mathematically rigorous
way of addressing these issues.  The classical configurations
in the universal path integral are configurations of discrete sequences 
of symbols such as integers or bits, with actions that are computable
functions of those configurations.  The path integral sums over
all possible computable configurations: as a result, any path
integral that is itself computable -- for example, a conventional path integral
performed in lattice gauge theory -- is contained
as a sub-integral in the universal path integral.  As will be shown, 
the universal path integral gives well-defined predictions
for probabilities of events or sequences of events. 
In particular, the universal path integral provides a mechanism
for how the classical world that we see around us arises out
of constructive and destructive interference between different
quantum paths.

\bigskip\noindent{\it The universal path integral}

Now define the universal path integral.
The method of definition relies on the theory of algorithmic
probability [5-8].  
Let $U$ represent the operation of a classical universal
computer that takes input programs bit strings $b=b_1 b_2 \ldots$
written in language $L$, producing as output the bit string $U(b)$.
The bit strings can be either finite or infinite in length.
We can think of such bit strings as real numbers in the
interval $[0,1]$, written in binary, e.g., $0.1011 = 11/16$.
If $U$ halts in finite time given input $b$, then it reads only a finite
number of bits of $b$.  Call such a set of 
bits a halting program $p$: all input bit strings that begin with the 
same program $p$ yield the same output $U(p)$. 
That is, programs are `prefix-free': 
no halting program is the prefix of another.
A halting program $p$ then corresponds to a sub-interval
of measure $2^{-|p|}$ where $|p|$ is the length of the program $p$:
if one generates the bits of the input program by flipping a fair
coin, the probability that the first $|p|$ bits of the input
string yield the program $p$ is $2^{-|p|}$.

Because $U$ is universal, some programs never halt and give an output.
Membership in the set of halting programs cannot be computed [5-7] (the halting
problem).  To cope with the halting problem, introduce a computer that
always halts in finite time $t$:
$U_t(b) =  U(b) $ if $U$ halts in time $t$ given input $b$, $U_t(b) = 0$, 
otherwise. 
The universal path integral is then
$$  \Sigma_L = \lim_{t\rightarrow\infty} \int_0^1  e^{2\pi i U_t(b)} db. \eqno(2)$$
Because programs are prefix free, the universal path
integral can be written
as a universal path sum over programs for $U_t$, all of which now halt:
$$\Sigma_L = \lim_{t\rightarrow\infty}  
\sum_{p: |p| \leq t} 2^{-|p|} e^{i2\pi U_t(p)}.
\eqno(3)$$
The universal path integral $\Sigma_L$ is a complex number with
amplitude between zero and one.

Because the halting set is uncomputable, the $t\rightarrow\infty$ limit
in the universal path integral/sum converges more slowly than any 
computable sum.  The amplitude for everything cannot be computed 
in practice.  While it might sound bad at first, the uncomputability of
$\Sigma$ is in fact acceptable: we are interested not in the absolute
amplitude for everything, but rather in the amplitudes
that predict the results of experiments, given our observations.
As will be seen below, such amplitudes represent computable 
sub-integrals of $\Sigma$. 

To make the connection to conventional path integrals, we can
assume that programs written in language $L$ 
consist of two parts, $p=p_0p_1$, where the first part
of the program $p_0$ specifies a finite set of paths $W(p_0)$ 
and the method for computing an action for members of
that set  $w \in W(p_0)$.  The second part of the
program $p_1$ picks out a particular path $w(p_0p_1) = w(p)$ 
from the set, and generates the action $U(p)$ for that path.

The universal path integral sums over
all computable configurations of bits and actions.
It includes as sub-integrals all the path integrals that
physicists would like to perform.  Even if the desired path integral
consists of a continuous, uncountable set of paths, the universal
path integral nonetheless contains sub-integrals that approximate
that integral to any desired accuracy. For example,
any lattice gauge theory path integral
over a finite lattice with the field values truncated to
a finite precision corresponds to
some finite interval of the universal path integral.

\bigskip\noindent{\it Quantum computing and the universal path integral}

All quantum computations are included in the universal path integral.
A quantum computation can always be written as a sequence of
controlled-NOT gates, Hadamard gates, and so-called $\pi/8$
gates (rotations by $\pi/4$ about the $z$-axis) [8].  Controlled-NOT
gates flip quantum bits conditioned on the values of other bits, 
Hadamard gates take qubits to equal superpositions of $|0\rangle$
and $|1\rangle$, and $\pi/8$ gates apply a phase of $e^{i\pi/8}$
to $|0\rangle$ and $e^{-i\pi/8}$ to $|1\rangle$.  A quantum 
computation can then be written as a uniform superposition of
sequences of bit configurations (paths)
determined by the quantum logic gates of the
computation, where each Hadamard gate doubles the number of paths
and each $\pi/8$ gate applies a phase to each path. 

Conversely, 
the entire universal path integral can be computed on a quantum
computer in the infinite time limit.  The universal path integral 
is simply equal to 
$$\Sigma_L =  \lim_{t\rightarrow\infty} 
\langle \Psi |  V^t_L  |\Psi\rangle,\eqno(4)$$
where $|\Psi\rangle = \int_0^1 |b\rangle db$ is the uniform
sum of all input bit strings, and $V^t_L$ is the action of the quantum computer
that implements the transformation 
$V^t_L |p\rangle = e^{2\pi i U_t(p)} |p\rangle$.  
That is, the amplitude modulus squared of the universal path integral 
$\Sigma_L$ is the probability that the quantum computer remains in its initial
state.  The real and imaginary parts of the amplitude can be extracted
in a similar fashion.  For example, start in the initial
state $ |\Psi\rangle \otimes (1/\sqrt 2)( |0\rangle + |1\rangle)$,
and use the quantum computer to construct the state
$1/\sqrt 2( V^t_L  |\Psi\rangle|0\rangle +
\bar V^t_L  |\Psi\rangle|1\rangle$.  The probability that
the quantum computer remains in its initial state is then proportional
to the square of the real part of the amplitude.   Sub-integrals
of the universal path integral are computable by a quantum
computer in an analogous fashion.

\bigskip\noindent{\it Obtaining probabilities from amplitudes}

Let's see how the universal path integral assigns probabilities
to events.  Use the two-step description $p=p_0p_1$ given above,
where $p_0$ picks out a set of paths $W(p_0)$ and $p_1$ picks out 
a particular path $w(p)$ within that set.  A coarse-grained
path $\tilde w$ is some subset of $W$.  For example, if $W$ is a set of 
bit strings, $\tilde w$ could be the set of bit strings where the 
first bit takes on the value 0.  The amplitude for $\tilde w$ is 
$${\cal A}(\tilde w) = 
\sum_{p:w(p) \in \tilde w } 2^{-|p|} e^{2\pi i U(p)}.\eqno(4)$$

The probability for an event in quantum mechanics is usually
taken to be proportional to the square of the magnitude of the amplitude
for the event:  
$$p(\tilde w)  \propto {| {\cal A}(\tilde w) |^2 }.\eqno(5)$$
We must be careful here, as the prescription that probability
is proportional to amplitude squared conventionally
refers to probabilities for the outcomes of measurements.
It is not yet clear what a `measurement' consists of here:
any measurement apparatus must itself be somehow contained within
the path integral.  Fortunately, the method of consistent or decoherent
histories gives us a well-established way to proceed [9-14]: 
even in the absence
of measurement apparatus, we can still assign probabilities as
long as the probability sum rules are obeyed.  

Consider two
non-overlapping coarse-grained states $\tilde w$ and $\tilde w'$,
e.g., $\tilde w' = NOT ~ \tilde w$, the set complementary to $\tilde w$.
We would like to assign to these states the `probabilities'
$$ p(\tilde w) = {| {\cal A}(\tilde w) |^2}, \quad 
p(\tilde w') =  {| {\cal A}(\tilde w') |^2}.\eqno(6)$$
Since $\tilde w$, $\tilde w'$ correspond to mutually exclusive
sets of events, to qualify as probabilities $ p(\tilde w),  p(\tilde w')$ 
should satisfy the probability sum rule:
$$p( \tilde w ~ OR~ \tilde w') =  p(\tilde w) +  p(\tilde w').\eqno(7)$$
In other words, defining the amplitude  
$${\cal A}(\tilde w ~OR ~ \tilde w') = 
\sum_{w\in  \tilde w \cup \tilde w'} {\cal A}(w),\eqno(8)$$
we require that
$$ | {\cal A}(\tilde w ~OR ~ \tilde w') |^2 
= | {\cal A}(\tilde w) |^2 + | {\cal A}(\tilde w')|^2.\eqno(9)$$
This requirement to obey the probability sum rule is equivalent
to demanding that
$$ {\rm Re}  {\cal A}(\tilde w) \bar {\cal A}(\tilde w') = 0.\eqno(10)$$

More generally, two non-overlapping coarse-grained states obey
the probability sum rule to accuracy $\epsilon$ if
$$ { ({\rm Re}{\cal A}(\tilde w) \bar {\cal A}(\tilde w'))^2
\over | {\cal A}(\tilde w) |^2 | {\cal A}(\tilde w')|^2} \leq \epsilon.
\eqno(11)$$
This is the usual requirement for destructive interference
from conventional quantum mechanics: an observer trying to
determine whether the sequences of events $\tilde w$ and $\tilde w'$
obey the probability sum rule would have to repeat the
experiment of seeing how many times $\tilde w$ or
$\tilde w'$ occured $O(1/\epsilon^2)$ times in order
to discern deviations from the probability sum rule
due to the effects of quantum interference.  For
example, if $\tilde w$ represents a particle showing
up in one region of the screen in a double slit experiment,
and if $\tilde w'$ represents a particle showing up in 
a non-overlapping part of the screen, the particle must
be sent through the slits $O(1/\epsilon^2)$ times to detect
the effects of quantum interference.

The method for obtaining probabilities for events from the 
path integral reveals a crucial difference between quantum
and classical algorithmic descriptions of the universe [15-17].
The classical algorithmic description simply assigns
algorithmic probabilities $2^{-|p|}$ to programs
and to the bit strings that they create.  In the classical case,
all computable structures are represented, but they do not interfere
with each other.  In the quantum case, by contrast, the world that
we see around is arises out of the interference
between different computable structures [16].  
Computable structures effectively conspire with each other
via constructive and destructive interference to create
the observable world.

\bigskip\noindent{\it Observable probabilities are 
language independent}

The universal path integral allows us to assign probabilities
to coarse grained histories that obey the probability sum
rules.  As defined so far, these probabilities depend
on the language $L$ used to define the universal
path integral.  The absolute probabilities defined
so far are not the same as the probabilities as measured
by observers `living in' the path integral, however: the
measured probabilities
are not absolute but are conditioned on the fact of the
observer's existence, and on the dynamics of the sub-integral
that the observer inhabits.  We now show the probabilities
for observables as measured by observers
are in fact independent of the $L$.

First of all, note that the universal path integral defined
according to one language contains 
the universal path integrals defined according
to every other language as a non-zero measure sub-integrals. 
For any two fully recursive languages $L$, $L'$,
there is a program $p_{L\rightarrow L'}$ written in $L$,
that instructs the computer to interpret what follows as
a series of instructions in language $L'$, where the symbols
recognized by $L'$ have been suitably encoded as bit strings
of finite length.  The universal path integral $\Sigma_{L'}$ is
then equal to the universal path integral $\Sigma_L$
restricted to the interval in which the
initial $|p_{L\rightarrow L'}|$ bits of the input program
in $\Sigma_L$ are $p_{L\rightarrow L'}$.   That is $\Sigma_L$
contains $\Sigma_{L'}$ as a sub-integral.
Note that this means that $\Sigma_L$ also contains itself
as a sub-integral (and does so an infinite number of times).
It also means that the amplitudes for events determined
by the universal path integral defined by $L'$ differ
by at most a multiplicative constant from those defined
by $L$.

Now consider the probabilities for events as observed by
someone `living' in the path integral.   The `life' of such an observer
is nothing more or less than a coarse-grained set of events,
whose joint probabilities are those for a system that is
gathering and processing information about other systems
to which it has access.  In Gell-Mann's nomenclature [11],
an observer is an information gathering and using system,
or IGUS, embedded in the path integral.
This observer sees events occurring with 
probabilities that depend on what part of the path integral
it occupies.  The key point is that the
observer has no access to the absolute probabilities of
events: it only has access to the probabilities of
events conditioned on its existence and on the laws
in its sector of the path integral.  In other words,
this observer, like all the rest of us,  
is subject to the weak anthropic principle: we
only have access to the part of the universe that supports
our existence.  

The universal path integral allows us to make this conditional
nature of probabilities precise: every observer occupies a
sub-integral of the path integral governed at bottom by a particular
language $L$, which is typically only partly known to that observer.
For example, in our case, the language $L$ specifies
the part of the path integral that governs the
laws of elementary particles and quantum gravity as they
figure in our particular sub-integral, acting
in a spacetime that obeys a particular set of initial
conditions.  Within the part of the path
integral specified by $L$, further sub-integrals pick
out the quantum accidents that specify our
observed laws of chemistry, biology, economics, etc. [11].
That is, $L$ is the `language of nature' for the part
of the universal path integral accessible to us as observers.
The theory of algorithmic inference [5-7] then implies
that Bayesian updating of the probabilities of theories 
based on repeated observation, will lead us closer and
closer to a full knowledge of the language of nature that specifies
our part of the path integral.  The nested nature of
the universal path integral implies that knowledge
of our `local' language of nature is just as good
as knowledge of the `global' language.

\bigskip\noindent{\it Discussion}

The universal path integral contains as sub-integrals all
computable path integrals, including 
the path integrals for lattice gauge theories.  
It supports quantum computation and can be computed by a quantum
computer in the infinite time limit.  It supplies quantum amplitudes
for fine-grained sequences of events, and predicts probabilities for
coarse-grained events.  The probabilities for events as viewed
by observers `living in' the universal path integral are independent of
the language by which the path integral is defined.
The world that we see around us arises out of quantum
interference between all possible computable structures.

As in all theories where the observed laws of Nature are
just some instance of possible laws, e.g., the string theory
landscape, the universal path integral theory of nature
is not as satisfying as a theory that predicts
the exact laws that we see from one fundamental principle.
Because of its intrinsic connection to Occam's razor,
however, the universal path integral does not discourage
us from looking for ever simpler laws: on the contrary,
it exhorts us to carry on in the search for simplicity,
in the hope that we will discover new regularities in
the language of Nature, and to use those regularities to predict the
results of future observations.

\vfill

\noindent{\it Acknowledgements:} This work was supported by
the W.M. Keck Center for Extreme Quantum Information Theory (xQIT),
DARPA, ARO under a MURI program, NSF, ENI via the MIT Energy Initiative,
Lockheed Martin, Intel, Jeffrey Epstein, and by FQXi.  The authors would like to
thank Janna Levin and Max Tegmark for helpful discussions.

\vfil\eject

\noindent{\it References}

\bigskip
\noindent{1.} L. Susskind, `The anthropic landscape of string
theory,' in B. Carr, {\it Universe or multiverse},
Cambridge University Press, 2007; arXiv:hep-th/0302219.

\bigskip\noindent{2.} K.G. Wilson, J. Kogut, {\it Rev. Mod. Phys.}
{\bf 55}, 775 (1983).

\bigskip\noindent{3.} H.J. Rothe, {\it Lattice Gauge Theories:
an Introduction}, World Scientific, Singapore (2005).

\bigskip\noindent{4.} A. Linde, {\it J. Cosm. Astro. Phys.} {\bf 22},
0701 (2007).

\bigskip\noindent{5.} R.J. Solomonoff, 
{\it Information and Control} {\bf 7}, 1-22 (1964).

\bigskip\noindent{6.}
G.J. Chaitin, {\it  Algorithmic Information Theory,}
Cambridge University Press, Cambridge, (1987).

\bigskip\noindent{7.}
A.N. Kolmogorov, 
{\it Problems of Information Transmission} {\bf 1},
1-11, (1965).

\bigskip\noindent{8.} M.A. Nielsen, I.L. Chuang,
{\it Quantum Computation and Quantum Information},
Cambridge University Press, Cambridge (2000).






\bigskip\noindent{9.} 
R. Griffiths, {\it J. Stat. Phys.} {\bf 36}, 219 (1984).

\bigskip\noindent{10.}
R. Omn\'es, {\it J. Stat. Phys.} {\bf 53}, 893, 933, 957 (1988);
 {\it J. Stat. Phys.} {\bf 57}, 359 (1989); {\it Rev. Mod. Phys.}
{\bf 64}, 339 (1992); {\it The Interpretation of Quantum
Mechanics}, Princeton University Press, Princeton, 1994.

\bigskip\noindent{11.} 
M. Gell-Mann, J.B. Hartle
{\it Phys. Rev. D} {\bf 47}, 3345-3382 (1993); gr-qc/9210010.

\bigskip\noindent{12.} J.J. Halliwell, {\it Phys. Rev. D}
{\bf 58} 105015 (1998), quant-ph/9805062;
{\it Phys. Rev. D} {\bf 60} 105031 (1999), quant-ph/9902008;
{\it Phys. Rev. Lett.} {\bf 83} 2481 (1999), quant-ph/9905094;
`Decoherent Histories for Spacetime Domains,' in
{\it Time in Quantum Mechanics,} edited by J.G.Muga, R. Sala Mayato
and I.L.Egususquiza (Springer, Berlin, 2001), quant-ph/0101099.

\bigskip\noindent{13.} J.B. Hartle,  {\it Phys. Scripta T} {\bf 76}, 67 (1998),
gr-qc/9712001.

\bigskip\noindent{14.} F. Dowker and A. Kent, {\it Phys. Rev. Lett.} {\bf 75},
3038 (1995).

\bigskip\noindent{15.} M. Tegmark, {\it Annals of Physics} {\bf 270}, 
1-51 (1998); arXiv/gr-qc/9704009.

\bigskip\noindent{16.}  S. Lloyd,  
{\it Complexity}, {\bf3/1}, 32-35 (1997); arXiv:quant-ph/9912088.

\bigskip\noindent{17.} J. Schmidhuber, in 
{\it Foundations of Computer Science: Potential - Theory - Cognition,}
Lecture Notes in Computer Science, C. Freksa, ed. 201-208, Springer, (1997);
arXiv:quant-ph/9904050.

\vfill\eject\end